\documentclass[conference]{IEEEtran}

\pdfoutput=1

\newlength{\figwidth}
\usepackage{xcolor}
\definecolor{links}{rgb}{0,0,0}   
\definecolor{urls}{rgb}{0,0,0.8}    
\definecolor{cites}{rgb}{0,0,0.8}   

\usepackage[nosort]{cite}        
\usepackage{url} 
\usepackage[intlimits]{amsmath}
\usepackage{bbm}
\usepackage{graphicx} 
\usepackage{paralist}
\usepackage{fancyref}
\usepackage[stretch=16,shrink=16,step=4]{microtype}
\usepackage{vmr-symbols-vecbold}
\usepackage{standard-macros}
\usepackage{mathbbol}
\usepackage{float}
\usepackage[acronym]{glossaries}

\DeclareSymbolFontAlphabet{\amsmathbb}{AMSb}%

\newcommand{\lro}[1]{\lefto({#1}\right)}																
\newcommand{\lrbo}[1]{\lefto \lbrace {#1} \right \rbrace}															
\newcommand{\lrho}[1]{\lefto [ {#1} \right ]}																				

\newcommand{\lr}[1]{\left({#1}\right)}																
\newcommand{\lrb}[1]{\left \lbrace {#1} \right \rbrace}															
\newcommand{\lrh}[1]{\left [ {#1} \right ]}																				

\safemath{\dopplerspread}{B_D}																								
\safemath{\delayspread}{T_D}																									
\safemath{\ncoh}{n\sub{c}}																										
\safemath{\ntx}{n\sub{t}} 																											
\safemath{\nrx}{n\sub{r}}																											
\safemath{\ntxt}{\tilde{n\sub{t}}}																											
\safemath{\cb}{\ensuremath{L}} 																								
\safemath{\cl}{\ensuremath{n}} 																								
\safemath{\txanto}{{\ensuremath{\tilde{m}_t}}} 																		
\safemath{\cs}{M} 																														
\safemath{\idPustm}{\ensuremath{S_{k}}}
\safemath{\error}{\ensuremath{\epsilon}} 																				
\safemath{\eexp}{\ensuremath{\mathcal{E}}} 																			
\safemath{\nsubc}{n\sub{s}}			 																						
\safemath{\nofdm}{n\sub{o}} 																									
\safemath{\bc}{\ensuremath{B_c}} 																							
\safemath{\ts}{\ensuremath{T_s}} 																							
\safemath{\nrb}{\ensuremath{n_{rb}}} 																						
\safemath{\nres}{\ell}
\newcommand{\cgauss}[2]{\mathcal{CN}\lro{ \ensuremath{#1, #2}  }}   								
\safemath{\maxk}{M^*\lr{\nres, \nsubc, \nofdm, \epsilon, \rho}}
\safemath{\Rmax}{R^*}
\safemath{\np}{n\sub{p}}

\safemath{\randmatA}{\amsmathbb{A}}
\safemath{\randmatB}{\amsmathbb{B}}
\safemath{\randmatC}{\amsmathbb{C}}
\safemath{\randmatD}{\amsmathbb{D}}
\safemath{\randmatE}{\amsmathbb{E}}
\safemath{\randmatF}{\amsmathbb{F}}
\safemath{\randmatG}{\amsmathbb{G}}
\safemath{\randmatH}{\amsmathbb{H}}
\safemath{\randmatI}{\amsmathbb{I}}
\safemath{\randmatJ}{\amsmathbb{J}}
\safemath{\randmatK}{\amsmathbb{K}}
\safemath{\randmatL}{\amsmathbb{L}}
\safemath{\randmatM}{\amsmathbb{M}}
\safemath{\randmatN}{\amsmathbb{N}}
\safemath{\randmatO}{\amsmathbb{O}}
\safemath{\randmatP}{\amsmathbb{P}}
\safemath{\randmatQ}{\amsmathbb{Q}}
\safemath{\randmatR}{\amsmathbb{R}}
\safemath{\randmatS}{\amsmathbb{S}}
\safemath{\randmatT}{\amsmathbb{T}}
\safemath{\randmatU}{\amsmathbb{U}}
\safemath{\randmatV}{\amsmathbb{V}}
\safemath{\randmatW}{\amsmathbb{W}}
\safemath{\randmatX}{\amsmathbb{X}}
\safemath{\randmatY}{\amsmathbb{Y}}
\safemath{\randmatZ}{\amsmathbb{Z}}

\safemath{\matSigma}{\bm{\Sigma}}
\safemath{\matPhi}{\bm{\Phi}}
\safemath{\matLambda}{\bm{\Lambda}}

\safemath{\randmatSigma}{\mathbb{\Sigma}}
\safemath{\randmatPhi}{\mathbb{\Phi}}
\safemath{\randmatLambda}{\mathbb{\Lambda}}

\usepackage{pgfplots}
\usepgfplotslibrary{external}
\tikzset{external/system call={latex \tikzexternalcheckshellescape -halt-on-error
-interaction=batchmode -jobname "\image" "\texsource" &&
dvips -o "\image".ps "\image".dvi}}
\makeatletter
\def\@IEEEinterspaceratioM{0.265}
\def\@IEEEinterspaceMINratioM{0.1651}
\def\@IEEEinterspaceMAXratioM{0.38}

\def\@IEEEinterspaceratioB{0.31}
\def\@IEEEinterspaceMINratioB{0.19}
\def\@IEEEinterspaceMAXratioB{0.38}
\@IEEEtunefonts
\makeatother
\newcommand\marksize{1.5} 
\newcommand\linew{1pt} 

\makeglossaries

\newacronym{lte}{LTE}{\emph{long term evolution}}
\newacronym{ue}{UE}{user equipment}
\newacronym{ul}{UL}{uplink}
\newacronym{dl}{DL}{downlink}
\newacronym{3gpp}{3GPP}{\emph{3rd generation partnership project}}
\newacronym{rb}{RB}{resource block}
\newacronym{tti}{TTI}{transmission time interval}
\newacronym{ofdm}{OFDM}{orthogonal frequency-division multiplexing}
\newacronym{iid}{i.i.d.}{identical and independently distributed}
\newacronym{psd}{PSD}{power spectral density}
\newacronym{mimo}{MIMO}{multiple-input multiple-output}
\newacronym{bler}{BLER}{block error probability}
\newacronym{mtc}{MTC}{machine-type communication}
\newacronym{csi}{CSI}{channel state information}
\newacronym{ustm}{USTM}{\emph{unitary space-time modulation}}
\begin{document}

\IEEEoverridecommandlockouts

\title{Low-latency Ultra-Reliable 5G Communications: Finite-Blocklength Bounds and Coding Schemes}
%
%

%
 \author{\IEEEauthorblockN{Johan \"Ostman$^1$, Giuseppe Durisi$^1$, Erik G. Str\"om$^1$, Jingya Li$^2$, Henrik Sahlin$^2$, and Gianluigi Liva$^3$\\
 $^1$Chalmers University of Technology,  Gothenburg, Sweden; $^2$Ericsson Research, Gothenburg, Sweden;\\ $^3$Deutsches Zentrum f\"ur Luft- und Raumfahrt (DLR), Wessling, Germany}
 \thanks{Accepted for publication in the 2017 IEEE Conference on Systems, Communications and Coding.}}

%
%
\maketitle

\begin{abstract}

Future autonomous systems require wireless connectivity able to support extremely stringent requirements on both latency and reliability.
In this paper, we leverage recent developments in the field of finite-blocklength information theory to illustrate how to optimally design wireless systems in the presence of such stringent constraints.
Focusing on a multi-antenna Rayleigh block-fading channel, we obtain bounds on the maximum number of bits that can be transmitted within given bandwidth, latency, and reliability constraints, using an orthogonal frequency-division multiplexing system similar to LTE. 
These bounds unveil the fundamental interplay between latency, bandwidth, rate, and reliability. 
Furthermore, they suggest how to optimally use the available spatial and frequency diversity.
Finally, we use our bounds to benchmark the performance of an actual coding scheme involving the transmission of short packets.

\end{abstract}

\section{Introduction} 
\label{sec:introduction}
The next generation of wireless cellular systems (5G) is expected to be a key enabler of future autonomous systems, be them connected vehicles, smart meters, or automated factories~\cite{osseiran14-05a,durisi16-01a}. 
The characteristics of the wireless data traffic typically generated within these autonomous systems is, however, drastically different from the one encountered in traditional broadband wireless applications: short data packets (on the order of hundreds of bits) that need to be delivered with stringent requirements in terms of latency and reliability. 

For example, \gls{mtc} for factory automation may involve the transmission of packets containing $100$~information bits within $100$~$\mu$s and with packet error probability not exceeding $10^{-9}$~\cite{johansson15-06a,yilmaz15-06a}. 
In traffic safety applications, one may need the packet error probability not to exceed $10^{-5}$~\cite{metis-project-deliverable-d1.113-04a}. 
These requirements are much more stringent than the ones that current wireless cellular systems, i.e., \gls{lte}, need to handle. 
Standardization activities are currently ongoing within the \gls{3gpp}, with the aim of evolving \gls{lte} and achieving these new requirements. 

One way to reduce latency is to assign to each user a \gls{rb} consisting of a smaller number of \gls{ofdm} symbols than currently done in \gls{lte}.\footnotemark{}  
This yields a shorter \gls{tti}. 
The impact of a reduced \gls{tti} on the performance of \gls{lte} has been recently analyzed in~\cite{ericsson16-04a,ericsson16-06a}. 

In order to increase reliability, one can use the available transmit and receive antennas to provide spatial diversity rather than spatial multiplexing. 
This has been investigated in~\cite{johansson15-06a,yilmaz15-06a} in a factory-automation scenario, under the assumption that perfect \gls{csi} is available at the receiver.
\footnotetext{In \gls{lte} release 13, an \gls{rb} comprises $12$ adjacent subcarriers over $7$ consecutive \gls{ofdm} symbol durations; in this paper, however, we allow an \gls{rb} to span an arbitrary number of adjacent subcarriers and consecutive \gls{ofdm} symbols.}

The problem of optimally designing a communication system operating under a stringent latency constraint can be addressed in a fundamental fashion using the finite-blocklength information theoretic bounds recently developed by Polyanskiy \emph{et al.}~\cite{polyanskiy10}. 
Using these tools, Durisi \emph{et al.}~\cite{durisi16-02a} developed bounds on the maximum coding rate over multi-antenna Rayleigh block-fading channels. 
Since these bounds do not assume the \emph{a priori} availability of perfect \gls{csi}, they unveil the fundamental tradeoff between exploiting spatial and time-frequency diversity (to obtain high reliability) on the one hand, and reducing channel-estimation overhead on the other hand. 
The bounds in~\cite{durisi16-02a}, however, require Monte-Carlo simulations and are difficult to compute for packet error probabilities below~$10^{-6}$. 
An alternative approach to obtaining achievability bounds on the maximum coding rate is through random-coding error exponent analyses~\cite{gallager68}. 
The random coding error exponent of Rayleigh-fading channels for the case when no \gls{csi} is available at the receiver has been obtained in~\cite{faycal99} for the single-input single-output case.
However, no error exponent results are available for the no-\gls{csi} multiple-antenna case.

\paragraph*{Contribution}
In this paper, we analyze the problem of designing an \gls{ofdm} based system (such as \gls{lte}) able to satisfy a given set of requirements on reliability, latency, and bandwidth occupancy.
The specific contributions are as follows. 
We use the information theoretic bounds recently developed in~\cite{durisi16-02a} for the multiple-antenna Rayleigh block-fading channel to analyze the tradeoff between latency, bandwidth, and rate for the case when each transmit packet comprises a certain number of \gls{rb}s that are assumed to be orthogonal in time and frequency, and subject to independent fading.
Our analysis applies to both the \gls{ul}, where we assume a fixed average power per use of the channel in time, and to the \gls{dl}, where we assume instead a \gls{psd} constraint. 
The analysis is performed for a target packet error probability of $10^{-5}$.
To understand how to optimally use spatial and frequency diversity when the requirement on packet error probability is $10^{-9}$ or lower (ultra-reliable communications), we extend the error-exponent analysis in~\cite{faycal99} to the case of multiple-antenna systems and provide an upper bound on the error probability for the case when the input distribution is the so called \gls{ustm}~\cite{marzetta99}. Finally, we use our bounds to benchmark the performance of a coding scheme based on pilot transmission and convolutional encoding of the information bits.

\paragraph*{Notation}
  Uppercase letters such as $X$ denote scalar random variables and their realizations are written in lowercase, e.g., $x$. 
  We use two different fonts to write deterministic matrices (e.g., $\matX$) and random matrices (e.g., $\randmatX$). 
  The superscript $\herm{}$ denotes  Hermitian transposition and $\tr\lro{\cdot}$ and $\det\lro{\cdot}$ denote the trace and the determinant of a given matrix, respectively. 
  The identity matrix of size $a\times a$ is written as $\matI_{a}$. We denote by $\mathcal{V}\lro{\cdot} $ the Vandermonde determinant~\cite[p. 22]{couillet11}. 
  The distribution of a circularly symmetric complex Gaussian random variable with variance $\sigma^2$ is denoted by $\cgauss{0}{\sigma^2}$. 
  Finally, $\log\lro{\cdot}$ indicates the natural logarithm, $\lrh{a}^+$ stands for $\max\lrbo{0, a}$, $\lfloor \cdot \rfloor$ is the floor operator, $\Gamma\lro{\cdot}$ denotes the Gamma function, and $\Ex{}{\cdot}$ denotes the expectation operator.

\section{System Model} 
\label{sec:system_model}

\begin{figure}[t]
\centering
\def\svgwidth{0.9\columnwidth}
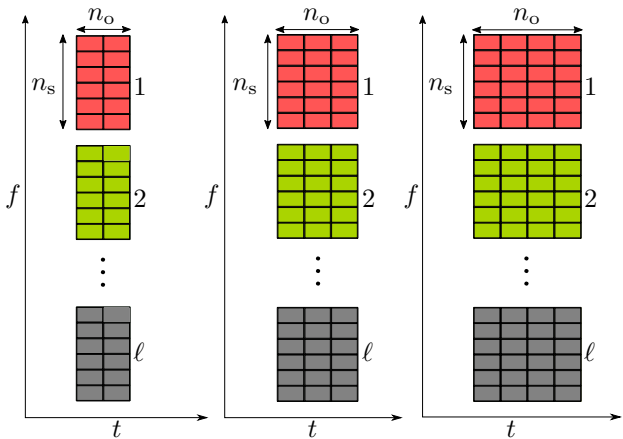
\caption{An example of three different UE resource allocations. Here, $\nsubc = 6$, $\nofdm = \lrb{2,3,4}$. The fading process is assumed constant over an \gls{rb} and the \gls{rb}s are assumed to fade independently (\gls{rb} spacing larger than the channel coherence bandwidth).}
\label{fig:ShortTTI}
\end{figure}

  We consider a wireless multiple-antenna communication system employing \gls{ofdm}, similar to what is used in \gls{lte}~\cite{dahlman11}. 
  As shown in Fig.~\ref{fig:ShortTTI}, a UE is assigned $\nres$ \gls{rb}s that are orthogonal in frequency, and constitute a packet.\footnotemark{}
  An \gls{rb} consists of $\nofdm$ OFDM symbols, each one spanning $\nsubc$ consecutive subcarriers. 
  Hence, an \gls{rb} contains a total of $\ncoh = \nofdm\nsubc$ time-frequency slots, also referred to as resource elements in \gls{lte}.
  Note that $\nofdm$ is related to the packet duration (in \gls{lte}, this quantity is referred to as \gls{tti}), whereas $\nsubc\nres$ is related to the bandwidth assigned to a given UE. 
  In \gls{lte}, the duration of an \gls{ofdm} symbol is approximately $71.4$ $\mu$s and the subcarrier spacing is $15$ kHz. Hence, an \gls{rb} consisting of $\nofdm = 7$ OFDM symbols and $\nsubc = 12$ subcarriers occupies $180$ kHz and lasts for $0.5$ ms. 
  Obviously, decreasing the number $\nofdm$ of \gls{ofdm} symbols within each \gls{rb} results in shorter delays. 
  This is currently under investigation within \gls{3gpp}~\cite{ericsson16-04a}.
  \footnotetext{In \gls{lte} \gls{ul}, the product $\nsubc\nres$ has to be a multiple of $2$, $3$ and $5$ due to implementation constraints. This will not be taken into account in this paper.}
  
We assume the channel fading to stay constant within each \gls{rb} and to change independently from \gls{rb} to \gls{rb} (block-fading model~\cite{marzetta99}). 
This assumption is reasonable for propagation environments characterized by low delay and Doppler spreads. 
One such example is the so called \gls{lte} pedestrian model, where the coherence bandwidth and the delay spread are approximately $23$~MHz and $200$~ms, respectively~\cite{etsi16}. 
The number of frequency diversity branches, i.e., the number of independent fading realizations within a given packet, is equal to the number of resource blocks $\nres$. 
We shall focus on Rayleigh fading. We shall also assume that the channels between each transmit-receive antenna pair fade independently (no spatial correlation).
    
   
The channel input-output relation within the $k$th \gls{rb}, for the case when the number of transmit antennas is $\ntx$ and the number of receive antennas is $\nrx$, can be expressed as:
   \begin{IEEEeqnarray}{rCl}
	\label{eq:ChannelModel}
	\randmatY_k = \matX_k \randmatH_k + \randmatW_k, \quad k=1,\dots, \nres.
	\end{IEEEeqnarray}
Here, $\matX_k\in \complexset^{\nofdm\nsubc \times \ntx}$ and $\randmatY_k \in \complexset^{\nofdm\nsubc \times \nrx}$ are the transmitted and received matrices, respectively; $\randmatH_k\in \complexset^{\ntx \times \nrx}$ is the fading matrix, whose entries are \gls{iid} $\cgauss{0}{1}$ random variables. Finally, $\randmatW_k\in \complexset^{\nofdm\nsubc \times \nrx}$, which denotes the thermal noise at the receiver, has independent $\cgauss{0}{1}$-distributed entries. The processes $\lrb{\randmatH_k}$ and $\lrb{\randmatW_k}$ are \gls{iid} across $k$ and are mutually independent. 

Throughout the paper, we shall assume that the realizations of the random fading matrices $\lrb{\randmatH_k}_{k=1}^{\nres}$ are unknown to the transmitter and the receiver. 
As discussed in, e.g., \cite{durisi16-02a}, \cite{yang13}, and \cite{lapidoth05}, this allows us to take into account the potential rate loss caused by the transmission of training sequences for channel estimation at the receiver.

Next, we define a channel code for the channel~(\ref{eq:ChannelModel}) using standard information-theoretic terminology (see, e.g.,~\cite{polyanskiy10}, \cite{durisi16-02a}).
\begin{dfn}
\label{dfn:ChannelCode}
An $\lr{\nres, \nsubc, \nofdm, M, \epsilon, \rho}$\textendash code for the channel~(\ref{eq:ChannelModel}) consists of
\begin{itemize}
\item An encoder $f:\lrb{1, \dots, M} \rightarrow \complexset^{\nofdm \nsubc \times \ntx \nres}$ that maps a message $J \in \lrb{1, \dots, M}$ to a codeword $\matC \lro{J} \in \lrb{\matC_1, \dots, \matC_M}$. Each codeword can be expressed as a concatenation of $\nres$  subcodewords, each spanning an \gls{rb}. Specifically, $\matC_m = \lrh{\matC_{m,1}, \dots, \matC_{m, \nres}}$, $m \in\lrb{1, \dots, M}$, where $\matC_{m,k}\in \complexset^{\nofdm\nsubc\times \ntx}$ for $k=1,\dots,\nres$. Each subcodeword satisfies the power constraint 
   \begin{IEEEeqnarray}{rCl}
	\label{eq:PowerConstraint}
	 \tr\lro{\herm{\matC_{m,k}} \matC_{m,k}}= \rho.
	\end{IEEEeqnarray}

\item A decoder $g: \complexset^{\nofdm \nsubc \times \nrx \nres} \rightarrow \lrb{1, \dots, M}$ that satisfies the maximum error probability constraint
   \begin{IEEEeqnarray}{rCl}
	\label{eq:Encoder}
	\max_{1\leq j \leq M} \mathrm{Pr}\lrbo{g\lro{\randmatY^{\nres}} \neq J \given J=j} \leq \epsilon
	\end{IEEEeqnarray}
	where $\randmatY^{\nres} = \lrh{\randmatY_1, \dots, \randmatY_\nres}$ is the channel output induced by codeword $\matX^{\nres} = \lrh{\matX_1,\dots , \matX_{\nres}}=f\lro{j}$ through~(\ref{eq:ChannelModel}).
\end{itemize}

\end{dfn}
  
 For the \gls{ul}, we shall set $\rho$ in~(\ref{eq:PowerConstraint}) as follows:
\begin{IEEEeqnarray}{rCl}
	\label{eq:SubPowerConstraintUL}
	\rho= \nofdm \rho\sub{u}/\nres.
\end{IEEEeqnarray}
Here, $\rho\sub{u}$ can be thought as the average SNR per use of the channel in time (recall that the noise is assumed to have unit variance). 
In the \gls{dl}, we shall instead assume a constraint on the \gls{psd}, i.e., on the average SNR per time-frequency slot. 
Specifically,
\begin{IEEEeqnarray}{rCl}
	\label{eq:SubPowerConstraintDL}
	\rho= \nofdm\nsubc \rho\sub{d}.
\end{IEEEeqnarray}
The subcodeword power constraints~(\ref{eq:SubPowerConstraintUL}) and~(\ref{eq:SubPowerConstraintDL}) imply the per-codeword power constraints $\tr\lro{\herm{\matC_{m}} \matC_{m}}= \nofdm  \rho\sub{u}$ and $\tr\lro{\herm{\matC_{m}} \matC_{m}}= \nofdm \nsubc \nres \rho\sub{d}$, $m=1,\dots, M$, for the \gls{ul} and the \gls{dl}, respectively. 
Constraint~(\ref{eq:SubPowerConstraintUL}) is motivated by the limited battery power at the UE, whereas constraint~(\ref{eq:SubPowerConstraintDL}) captures that cellular base-stations need to fulfill spectral transmission masks.

The \emph{maximum coding rate} $\Rmax$ denotes the largest number of bits per time-frequency slot that can be transmitted with probability of error no larger than $\epsilon$, for given $\rho$, $\nsubc$, $\nres$ and $\nofdm$: 
 \begin{IEEEeqnarray}{rCl}
\label{eq:MaxRate}	
	\Rmax \triangleq \sup\lrb{\frac{\log_2\lro{M}}{\nsubc\nofdm\nres} : \exists \lr{\nres, \nsubc, \nofdm, M, \epsilon, \rho} \textendash \text{code}}.
\end{IEEEeqnarray}
For a given subcarrier spacing and a given OFDM symbol duration, $\Rmax$ is related to the largest number of bits $\lfloor \nofdm\nsubc\nres\Rmax \rfloor$ that can be transmitted with reliability~($1-\epsilon$) through the channel~(\ref{eq:ChannelModel}) for given latency and bandwidth constraints. 

\section{Finite-Blocklength Bounds} 
\label{sec:finite_blocklength_bounds}

Finite-blocklength bounds for the multiple-antenna Rayleigh block-fading channel were recently proposed in~\cite{durisi16-02a}. 
Here, we will review these bound and adapt them to our setting (differently from~\cite{durisi16-02a}, we allow coding over frequency, which requires a different power normalization). 
%
%
The following definition will turn out useful.
  
\begin{dfn}
\label{dfn:def1}
Assume that $\ncoh=\nofdm\nsubc$ is larger than the total number of antennas, $\ntx+\nrx$. Let $\matSigma_k$ be an $\ncoh\times\ncoh$ diagonal matrix with positive diagonal entries.
Let $\xi$ be a positive real constant, $q = \min\lrbo{\ntx, \nrx}$ and $p=\max\lrbo{\ntx, \nrx}$. For $k=1,\dots, \nres$ we define the random variable
\begin{IEEEeqnarray}{rCl}
\label{eq:infodens}
	\idPustm\lro{\matSigma_k, \xi} =  c\lro{\matSigma_k}  - \tr\lro{\herm{\randmatZ_k}\randmatZ_k} - \log\lro{\psi\lro{\randmatLambda, \xi}}
\end{IEEEeqnarray}
where $\lrb{\randmatZ_k}_{k=1}^{\nres}$ are independent complex Gaussian $\nofdm\nsubc \times \nrx$ matrices with \gls{iid} $\cgauss{0}{1}$ entries and $\randmatLambda = \diag\lro{\Lambda_1, \dots, \Lambda_{\nrx}}$ is a diagonal matrix whose diagonal entries are the ordered eigenvalues of $\herm{\randmatZ}_k \matSigma_k \randmatZ_k$. The function, $ c \lro{\matSigma_k}$ is given as follows:

\begin{IEEEeqnarray}{rCl}
  c \lro{\matSigma_k} &=&\ntx\lro{\ncoh-\ntx}\log\lro{\frac{\rho}{\ntx}} - \nrx\log\lro{\det\lro{\matSigma_k}}   \nonumber \\
  &&-\:\ntx\lro{\ncoh-\ntx - \nrx}\log\lro{1+\frac{\rho}{\ntx}} \nonumber \\
  &&+\: \sum_{u=1}^{\ntx} \log\lro{\Gamma\lro{u}}   - \sum_{u=\ncoh-q+1}^{\ncoh} \log\lro{\Gamma\lro{u}}.
\end{IEEEeqnarray}
Furthermore,
\begin{IEEEeqnarray}{rCl}
	\psi\lro{\randmatLambda, \xi} &=& \frac{\det\lro{\matM\lro{\randmatLambda,\xi}}}{\mathcal{V}\lro{\randmatLambda}}  \prod_{i=1}^{\nrx} \frac{\exp\lro{-\Lambda_i/\lro{1+\rho/\ntx}}}{\Lambda_i^{ \ncoh-\nrx} }
\end{IEEEeqnarray}
where
\begin{IEEEeqnarray}{rCl}
    \label{eq:matM}
	\lrh{\matM\lro{\randmatLambda,\xi}}_{i,j} =
	\begin{cases}
			\Lambda_i^{\ntx-j}  \tilde{\gamma}\lro{\lrho{\ncoh+j-p-\ntx}^+, \Lambda_i \xi},  \\ 
			\quad  1\leq i\leq  \nrx,\quad 1\leq  j \leq \ntx; \\
\exp\lro{-\Lambda_i \xi}\lrho{\frac{\partial^{\ntx-j} }{\partial \delta^{\ntx-j}} \delta^{\ncoh-i}\big\vert_{\delta = \xi}}, \\
 			\quad  \nrx < i \leq p ,\quad 1< j\leq \ntx ;\\
   \Lambda_i^{\ncoh-j} \exp\lro{-\Lambda_i \xi},\\
    	\quad 1 \leq i\leq\nrx, \quad \ntx < j \leq p
	\end{cases} 
	\end{IEEEeqnarray}
	with 
\begin{IEEEeqnarray}{rCl}
\tilde{\gamma}\lro{n, x} \triangleq \frac{1}{\Gamma\lro{n}}\int_{0}^{x} t^{n-1}\exp\lro{-t}dt
\end{IEEEeqnarray}
denoting the regularized incomplete Gamma function. 
\end{dfn}  

With the help of Definition~\ref{dfn:def1}, we shall provide in the next two theorems an achievability (lower) and a converse (upper) bound on the maximum coding rate $\Rmax$ defined in (\ref{eq:MaxRate}).

\begin{thm}
\label{thm:DT}
The max. coding rate $\Rmax$ is lower-bounded as

\begin{IEEEeqnarray}{rCl}
\label{eq:DT}
	\Rmax  \geq \max \lrbo{\frac{\log_2\lro{M}}{\nsubc\nofdm \nres}: \epsilon_{\mathrm{ub}}\lro{M}\leq \epsilon}
\end{IEEEeqnarray}
where
\begin{IEEEeqnarray}{rCl}
	\epsilon_{\mathrm{ub}}\lro{M} =  \Ex{}{\exp\lro{-\lrho{\sum_{k=1}^{\nres} \idPustm\lro{\matSigma_k, \xi} - \log\lro{\cs-1}}^+}}. \nonumber \\
\end{IEEEeqnarray}
Here, $\idPustm\lro{\matSigma_k, \xi}$ is defined in~(\ref{eq:infodens}), $\xi=\rho/\lro{\ntx+\rho}$, and $\matSigma_k = \diag(\underbrace{\rho/\ntx+1, \dots, \rho/\ntx+1}_{\ntx}, \underbrace{1,\dots, 1}_{\ncoh-\ntx})$.
\begin{IEEEproof}
The bound is obtained by applying the dependence testing bound~\cite[Thm. 22]{polyanskiy10} to the channel~(\ref{eq:ChannelModel}) with input distribution chosen as \gls{ustm}. For details, see \cite[Thm. 1]{durisi16-02a}.
\end{IEEEproof}
\end{thm}

\begin{thm}
\label{thm:MC}
The max. coding rate $\Rmax$ is upper-bounded as
\begin{IEEEeqnarray}{rCl}
\label{eq:MC}
	&& \Rmax  \leq    \sup_{\lrbo{\matSigma_k}_{k=1}^{\nres}} \inf_{\gamma > 0}\frac{1}{\nsubc\nofdm \nres \log\lro{2}} \nonumber \\
  &&\times \: \lrbo{\gamma - \log\lro{\lrho{\mathrm{Pr}\lrbo{\sum_{k=1}^{\nres} \idPustm\lro{\tilde{\matSigma}_k, \xi} \leq \gamma} - \epsilon}^+}}. \nonumber \\*
\end{IEEEeqnarray}
Here, $\idPustm(\tilde{\matSigma}_k, \xi)$ is defined in~(\ref{eq:infodens}), $\xi=\rho/\lro{\ntx+\rho}$ and the matrices $\lbrace\tilde{\matSigma}_k\rbrace_{k=1}^{\nres}$ are given as follows
\begin{IEEEeqnarray}{rCl}
\tilde{\matSigma}_k = \begin{bmatrix} \matSigma_k + \matI_{\ntx} & 0 \\ 0 & \matI_{\ncoh-\ntx} \end{bmatrix}
\end{IEEEeqnarray}
with $\lrb{\matSigma_k}_{k=1}^{\nres}$ being $\ntx\times\ntx$ diagonal matrices with nonnegative elements satisfying the power constraint $\tr\lro{\matSigma_k} = \rho$.
\begin{IEEEproof}
 The proof relies on the metaconverse theorem~\cite[Thm.~28]{polyanskiy10}. The auxiliary distribution is chosen as the output distribution induced by an \gls{ustm} input through the channel~(\ref{eq:ChannelModel}). For details, see~\cite[Thm.~2 and Remark 2]{durisi16-02a}.
\end{IEEEproof}
 
\end{thm}

In the next section, the bounds in Theorem~\ref{thm:DT} and Theorem~\ref{thm:MC} will be used to characterize $\Rmax$ for given latency and bandwidth occupancy constraints. 
Our implementation of the numerical routines needed for the evaluation of these bounds (available as part of \texttt{spectre}\textendash\emph{short-packet communications toolbox}~\cite{durisi14-12b}) requires Monte-Carlo analysis, rendering these bounds difficult to compute for packet error probabilities below $10^{-5}$.
To address this problem, we shall complement these bounds with an achievability bound on $\Rmax$ based on Gallager's random coding error exponent that can be easily computed for low error probabilities.

\begin{thm}  
\label{thm:eexp}
Let $\ncoh=\nofdm\nsubc$ be larger than the total number of antennas $\ntx+\nrx$. Fix a rate $R$ and let $q=\min\lrbo{\ntx, \nrx}$. Let also $\randmatY = \randmatX\randmatH + \randmatW$ where $\randmatH$ and $\randmatW$ are defined as in~(\ref{eq:ChannelModel}) and $\randmatX =  (\rho / \ntx)\randmatPhi$, where $\randmatPhi \in \complexset^{\nofdm\nsubc \times \ntx}$ is unitary and isotropically distributed. Finally, let $\randmatLambda = \diag\lro{\Lambda_1, \dots, \Lambda_{\nrx}}$ denote the ordered eigenvalues of $\herm{\randmatY}\randmatY$.  
The average error probability $\bar{\epsilon}$ is upper-bounded by
\begin{IEEEeqnarray}{rCl}
\label{eq:eps_eexp}
\bar{\epsilon} \leq \min_{0\leq \mu\leq 1} \exp\lro{-\nres \lro{\mathcal{E}\lro{\mu} - \mu R}}
\end{IEEEeqnarray}
where
\begin{IEEEeqnarray}{rCl}
    \label{eq:eexp_rayl} 
     &&\mathcal{E}\lro{\mu} =c\lro{\mu} \nonumber \\
      && -\: \log \Ex{\randmatLambda}{ \lro{\frac{\prod_{i=1}^{{\nrx}}e^{\xi \Lambda_i}\Lambda_i^{\nrx-\ncoh}}  {\mathcal{V}\lro{\randmatLambda}} \det\lro{\matM\lro{\randmatLambda, \xi}}}^{\lro{1+\mu}}  } \IEEEeqnarraynumspace
\end{IEEEeqnarray}
with $\xi=\rho/\lro{\lro{1+\rho}\lro{1+\mu}}$ and
   \begin{IEEEeqnarray}{rCl}
    \label{eq:eexp_rayl_const}
   c\lro{\mu} &=& \lro{1+\mu} \log\lro{\frac{\lro{ 1+ \frac{\rho}{\ntx}}^{\frac{\nrx\ntx}{1+\mu}} \xi^{\ntx\lro{\ncoh-\ntx}}  \prod_{i=1}^{\ntx}\Gamma\lro{i} }{\prod_{i=\ncoh-q+1}^{\ncoh} \Gamma(i)} }. \nonumber \\
   \end{IEEEeqnarray}

 The matrix $\matM\lro{\randmatLambda, \xi}$ in~(\ref{eq:eexp_rayl}) is defined in~(\ref{eq:matM}). Furthermore, the probability distribution function of the ordered eigenvalues $\lr{\Lambda_1, \dots, \Lambda_{\nrx}}$ is given by
\begin{IEEEeqnarray}{rCl}
    \label{eq:eexp_rayl_pdflambda}
  f_{\randmatLambda}(\matLambda) = \frac{\exp\lro{-\sum_{i=1}^{\nrx} \lambda_i} \lro{\prod_{i=1}^{\nrx} \lambda_i} \mathcal{V}\lro{\matLambda}^2}{\prod_{i=1}^{\nrx} \Gamma\lro{\ncoh-i+1}\Gamma\lro{\nrx-i+1}}.
\end{IEEEeqnarray}

\begin{IEEEproof} 
This result follows essentially from \cite{faycal99} by choosing \gls{ustm} as input distribution. The details are omitted due to space constraints.
\end{IEEEproof}

\end{thm}

\begin{rem}
The average error probability $\bar{\epsilon}$ in~(\ref{eq:eps_eexp}) can be converted into maximum error probability (see (\ref{eq:Encoder})) by following a standard procedure (see, e.g.,  \cite[p. 204]{cover06}).
\end{rem}

Unfortunately, the expectation in Theorem~\ref{thm:eexp} seems formidable to solve in closed form. 
However, for small $\nrx$, say $\nrx \leq 3$, it can be efficiently evaluated numerically.
  

\section{Numerical Results} 
\label{sec:numerical_results}

In this section, we shall use the bounds~(\ref{eq:DT}),~(\ref{eq:MC}), and~(\ref{eq:eps_eexp}) to derive guidelines on the optimal design of the \gls{ofdm} system described in Section \ref{sec:system_model} as a function of the latency, bandwidth, and reliability constraints.

The numerical evaluation of the upper bound~(\ref{eq:MC}) is challenging because it requires one to maximize over the diagonal matrices $\lrb{\matSigma_k}_{k=1}^{\nres}$. 
Throughout this section we simplify the numerical evaluations by assuming $\matSigma_k = \lr{\rho/\ntx}\matI_{\ntx}$. 
The accuracy of this approximation was validated numerically in~\cite{durisi16-02a}.

\subsection{Dependency of $\Rmax$ on $\nres$ and $\nofdm$} 
\label{sec:_10_5_}

In this section, we shall use the bounds in Theorem~\ref{thm:DT} and~\ref{thm:MC} to investigate how $\Rmax$ depends on the number of resource blocks $\nres$ and the number of OFDM symbols $\nofdm$. 
We shall consider both a $1\times 2$ and a $2\times 2$ \gls{mimo} system in the \gls{ul} and both a $2\times 1$ and a $2\times 2$ \gls{mimo} system in the \gls{dl}. 
The target packet error probability is $10^{-5}$. 
Furthermore, we shall assume throughout this subsection that the number of subcarriers per \gls{rb}, $\nsubc$, is $12$ and consider both the case $\nofdm=2$ and $\nofdm=4$ OFDM symbols. 
For an OFDM symbol duration of $71.4$~$\mu$s (including cyclic prefix) as in LTE, these values of $\nofdm$ yield a packet duration of $142.8$~$\mu$s and $285.6$~$\mu$s, respectively. 
We shall also assume a subcarrier spacing of $15$~kHz (again as in \gls{lte}) so that we can relate the product $\nsubc \nres$ to the bandwidth assigned to a given UE.

Our results for the \gls{ul} are reported in Fig.~\ref{fig:UL_1x2} ($1\times 2$ \gls{mimo}) and Fig.~\ref{fig:UL_2x2} ($2\times 2$ \gls{mimo}). 
As expected, achievability and converse bounds are loose for small values of~$\nres$ and become progressively tighter as \nres (and, hence, the packet size $\nofdm\nsubc\nres$) increases. 
As expected, $\Rmax$ is larger for the case $\nofdm=4$ because the packet size is larger, which allows for more resilience against the additive noise. 
The crossing between the converse curve for the case $\nofdm=2$ and the one for the case $\nofdm=4$ for values of $\nres$ smaller than $3$ is merely a consequence of the looseness of our converse bound for very small values of $\nres$ (note that this crossing does not occur for the achievability bound). 
As far as the dependency of $\Rmax$ on~$\nres$ is concerned, we observe that our bounds are not monotonic in $\nres$, but that there exists an optimal $\nres$ (roughly about $\nres=5$ for the $\nofdm=4$ case) beyond which the maximum coding rate decreases. 
To the left of this optimal value, the main bottleneck is the limited time-frequency diversity, whereas to the right of this optimal value the main bottleneck is the low power per resource block available (the power scales inversely with $\nres$, see~(\ref{eq:SubPowerConstraintUL})).

In Fig.~\ref{fig:UL_2x2}, we consider the $2\times 2$ \gls{mimo} case. 
We see that adding a second transmit antenna is beneficial for small values of $\nres$. Indeed, for the case $\nofdm=4$, the achievable bound peaks at about $0.94$ bits per time-frequency slots compared to $0.54$ bits per time-frequency slots in the $1\times 2$ case. 
Furthermore, this peak occurs at smaller values of $\nres$ ($3$ instead of $5$), which implies that the additional spatial diversity provided by the second antenna reduces the need for frequency diversity.
This results in bandwidth savings. 
We note also that, as $\nres$ increases, the rate gains resulting from the use of a second antenna diminish. 
This is accordance to the well-known result that in the low-SNR regime, using a single antenna is optimal when the channel is not known to the receiver~\cite{verdu02}.

\begin{figure}[t]
\centering
\begin{tikzpicture}
\begin{axis}[%
xmin=1,
xmax=25,
ymin = 0,
xminorticks=true,
xlabel={Number of resource blocks $\nres$},
xtick={1,5,10,15,20,25},
xticklabels={$1$, $5$, $10$, $15$, $20$, $25$},
ylabel={Bits per time-frequency slot},
axis background/.style={fill=white},
grid=major,
legend pos= north west,
]
\addplot[name path = TTI2_MC, color=blue,solid,mark=*,mark options={solid, mark size=\marksize},line width=\linew] table [ y index={3}, col sep=comma] {../Data/R_2x1_TTI_2_eps_1e-05_T_24_DL.csv}coordinate[pos=0.80](ut1);\addlegendentry{Converse (\ref{eq:MC})}
\addplot[name path = TTI2_DT, color=red,solid,mark=square,mark options={solid, mark size=\marksize},line width=\linew] table [ y index={2}, col sep=comma] {../Data/R_2x1_TTI_2_eps_1e-05_T_24_DL.csv}coordinate[pos=0.80](pt1);\addlegendentry{Achievability (\ref{eq:DT})}
\addplot [color=gray, opacity=0.2, forget plot] fill between[of=TTI2_MC and TTI2_DT];

\coordinate (pt2) at ($(pt1) !.5! (ut1)$);
\draw (pt2) ellipse  (2pt and 4pt);
\coordinate (pt3) at ($(pt2)+ (0pt,-4pt)$);
\coordinate (pt4) at ($(pt3)+ (-5pt,-10pt)$);
\draw[<-] (pt3)--(pt4) node[anchor=north ] {$2\times 1$, $\nofdm=2$ ($142.8$ $\mu$s)};

\addplot[name path = TTI2_MC, color=blue,solid,mark=*,mark options={solid, mark size=\marksize},line width=\linew] table [ y index={3}, col sep=comma] {../Data/R_2x2_TTI_2_eps_1e-05_T_24_DL.csv}coordinate[pos=0.80](ut1);
\addplot[name path = TTI2_DT, color=red,solid,mark=square,mark options={solid, mark size=\marksize},line width=\linew, forget plot] table [ y index={2}, col sep=comma] {../Data/R_2x2_TTI_2_eps_1e-05_T_24_DL.csv}coordinate[pos=0.65](pt1);
\addplot [color=gray, opacity=0.2, forget plot] fill between[of=TTI2_MC and TTI2_DT];

\coordinate (pt2) at ($(pt1) !.5! (ut1)$);
\draw (pt2) ellipse  (2pt and 4pt);
\coordinate (pt3) at ($(pt2)+ (0pt,-4pt)$);
\coordinate (pt4) at ($(pt3)+ (+5pt,-10pt)$);
\draw[<-] (pt3)--(pt4) node[anchor=north] {$2\times 2$, $\nofdm=2$ ($142.8$ $\mu$s)};
\addplot[name path = TTI4_MC, color=blue,solid,mark=*,mark options={solid, mark size=\marksize},line width=\linew,dashed] table [ y index={3}, col sep=comma] {../Data/R_2x1_TTI_4_eps_1e-05_T_48_DL.csv}coordinate[pos=0.80](ut1);
\addplot[name path = TTI4_DT, color=red,solid,mark=square, mark options={solid, mark size=\marksize},line width=\linew, forget plot,dashed] table [y index={2}, col sep=comma] {../Data/R_2x1_TTI_4_eps_1e-05_T_48_DL.csv}coordinate[pos=0.60](pt1);
\addplot [color=gray, opacity=0.2, forget plot] fill between[of=TTI4_MC and TTI4_DT];

\coordinate (pt2) at ($(pt1) !.5! (ut1)$);
\draw (pt2) ellipse  (2pt and 3pt);
\coordinate (pt3) at ($(pt2)+ (0pt,+3pt)$);
\coordinate (pt4) at ($(pt3)+ (+5pt,+10pt)$);
\draw[<-] (pt3)--(pt4) node[anchor=south] {$2\times 1$, $\nofdm=4$ ($285.6$ $\mu$s)};

\addplot[name path = TTI4_MC, color=blue,solid,mark=*,mark options={solid, mark size=\marksize},line width=\linew,dashed] table [y index={3}, col sep=comma] {../Data/R_2x2_TTI_4_eps_1e-05_T_48_DL.csv}coordinate[pos=0.55](ut1);
\addplot[name path = TTI4_DT, color=red,solid,mark=square, mark options={solid, mark size=\marksize},line width=\linew, forget plot,dashed] table [y index={2}, col sep=comma] {../Data/R_2x2_TTI_4_eps_1e-05_T_48_DL.csv}coordinate[pos=0.7](pt1);
\addplot [color=gray, opacity=0.2, forget plot] fill between[of=TTI4_MC and TTI4_DT];

\coordinate (pt2) at ($(pt1) !.5! (ut1)$);
\draw (pt2) ellipse  (2pt and 4pt);
\coordinate (pt3) at ($(pt2)+ (0pt,+4pt)$);
\coordinate (pt4) at ($(pt3)+ (+5pt,+5pt)$);
\draw[<-] (pt3)--(pt4) node[anchor=south] {$2\times 2$, $\nofdm=4$ ($285.6$ $\mu$s)};

\addplot[draw=none]{0};

\end{axis}
\begin{axis}
[
	  xmin=1,
	  xmax=25,
      xlabel={Bandwidth [MHz]},
      xticklabels={{$0.18$}, {$0.9$}, {$1.8$}, {$2.7$}, {$3.6$}, {$4.5$}},
      xtick={1,5,10,15,20,25},
      hide y axis,
      axis x line*=top,
      xlabel near ticks
]
    \addplot[draw=none]{0};
    \end{axis}
\end{tikzpicture}
\caption{Achievability bound (\ref{eq:DT}) and converse bound (\ref{eq:MC}) on the maximum coding rate in a \gls{dl} $2\times 1$ and a \gls{dl} $2\times 2$ system for different number of resource blocks $\nres$. Here, $\rho\sub{d}=10$ dB, $\error = 10^{-5}$, and $\nsubc = 12$.}
\label{fig:DL_2x1}
\end{figure}
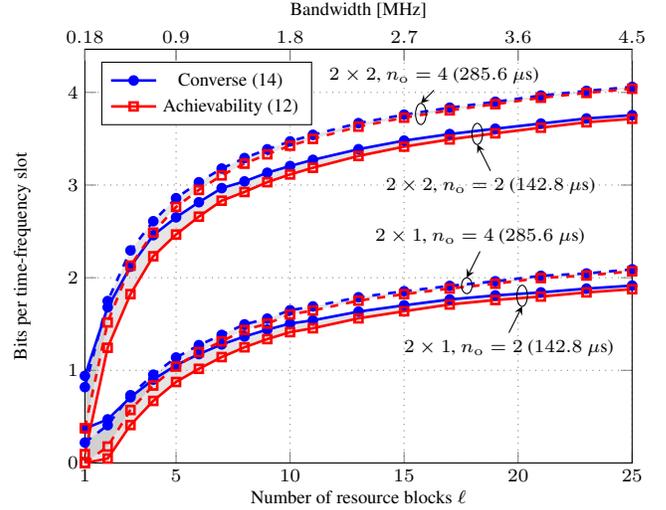

The \gls{dl} scenario is analyzed in Fig.~\ref{fig:DL_2x1} for the $2\times 1$ and $2\times 2$ cases. 
Since now the available power increases with $\nres$ because of the \gls{psd} constraint~(\ref{eq:SubPowerConstraintDL}), all curves become monotonic in $\nres$. 
As shown in the figures, our bounds allow one to estimate accurately the bandwidth and latency required to operate at a given rate. 
We see for example that for the $2\times 1$ \gls{mimo} case, one can operate at a rate of approximately $1.4$~bits per time-frequency slot using a packet of duration $285.6$~$\mu$s ($\nofdm=4$) and a bandwidth of $1.26$~MHz, or alternatively a packet of duration $142.8$~$\mu$s ($\nofdm=2$) and a bandwidth of $1.44$~MHz.

\begin{figure}[t]
\centering
\begin{tikzpicture}
\begin{axis}[%
xmin=1,
xmax=25,
ymin = 0,
ymax = 1.05,
xminorticks=true,
xlabel={Number of resource blocks $\nres$},
xtick={1,5,10,15,20,25},
xticklabels={$1$, $5$, $10$, $15$, $20$, $25$},
ylabel={Bits per time-frequency slot},
axis background/.style={fill=white},
grid=both,
legend pos= north east,
]
\addplot[name path = TTI2_MC, color=blue,solid,mark=*,mark options={solid, mark size=\marksize},line width=\linew] table [ y index={3}, col sep=comma] {../Data/R_1x2_TTI_2_eps_1e-05_T_24_UL.csv}coordinate[pos=0.70](ut1);\addlegendentry{Converse (\ref{eq:MC})}
\addplot[name path = TTI2_DT, color=red,solid,mark=square,mark options={solid, mark size=\marksize},line width=\linew] table [ y index={2}, col sep=comma] {../Data/R_1x2_TTI_2_eps_1e-05_T_24_UL.csv} coordinate[pos=0.70](pt1);\addlegendentry{Achievability (\ref{eq:DT})}
\addplot [color=gray, opacity=0.2, forget plot] fill between[of=TTI2_MC and TTI2_DT];

\coordinate (pt2) at ($(pt1) !.5! (ut1)$);
\draw (pt2) ellipse  (4pt and 6pt);
\coordinate (pt3) at ($(pt2)+ (0pt,-7pt)$);
\coordinate (pt4) at ($(pt3)+ (+10pt,-14pt)$);
\draw[<-] (pt3)--(pt4) node[anchor=north ] {$\nofdm=2$ ($142.8$ $\mu$s)};
 
\addplot[name path = TTI4_MC, color=blue,solid,mark=*,mark options={solid, mark size=\marksize},line width=\linew,dashed, forget plot] table [ y index={3}, col sep=comma] {../Data/R_1x2_TTI_4_eps_1e-05_T_48_UL.csv}coordinate[pos=0.80](ut1);
\addplot[name path = TTI4_DT, color=red,solid,mark=square, mark options={solid, mark size=\marksize},line width=\linew, forget plot,dashed] table [y index={2}, col sep=comma] {../Data/R_1x2_TTI_4_eps_1e-05_T_48_UL.csv}coordinate[pos=0.80](pt1);
\addplot [style={pattern=dots,pattern color=gray}] fill between[of=TTI4_MC and TTI4_DT];

\coordinate (pt2) at ($(pt1) !.5! (ut1)$);

\draw (pt2) ellipse  (2pt and 5pt);
\coordinate (pt3) at ($(pt2)+ (0pt,+5pt)$);
\coordinate (pt4) at ($(pt3)+ (+10pt,+10pt)$);
\draw[<-] (pt3)--(pt4) node[anchor=south] {$\nofdm=4$ ($285.6$ $\mu$s)};
 
\addplot[draw=none]{0};
\end{axis}

\begin{axis}
[
	  xmin=1,
	  xmax=25,
      xlabel={Bandwidth [MHz]},
      xticklabels={{$0.18$}, {$0.9$}, {$1.8$}, {$2.7$}, {$3.6$}, {$4.5$}},
      xtick={1,5,10,15,20,25},
      hide y axis,
      axis x line*=top,
      xlabel near ticks
]
    \addplot[draw=none]{0};
    \end{axis}

\end{tikzpicture}
\caption{Achievability bound (\ref{eq:DT}) and converse bound (\ref{eq:MC}) on the maximum coding rate in a UL $1\times 2 $ system for different number of resource blocks $\nres$. Here, $\rho\sub{u} = 20$ dB, $\error = 10^{-5}$, and $\nsubc= 12$. }
\label{fig:UL_1x2}
\end{figure}
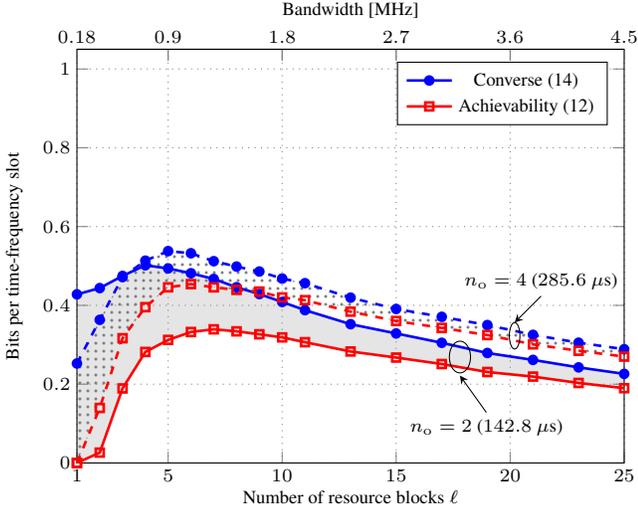

\begin{figure}[t]
\centering
\begin{tikzpicture}
\begin{axis}[%
xmin=1,
xmax=25,
ymin = 0,
ymax = 1.05,
xminorticks=true,
xlabel={Number of resource blocks $\nres$},
xtick={1,5,10,15,20,25},
xticklabels={$1$, $5$, $10$, $15$, $20$, $25$},
ylabel={Bits per time-frequency slot},
axis background/.style={fill=white},
grid=major,
legend pos= north east,
]
\addplot[name path = TTI2_MC, color=blue,solid,mark=*,mark options={solid, mark size=\marksize},line width=\linew] table [ y index={3}, col sep=comma] {../Data/R_2x2_TTI_2_eps_1e-05_T_24_UL.csv}coordinate[pos=0.80](ut1);\addlegendentry{Converse (\ref{eq:MC})}
\addplot[name path = TTI2_DT, color=red,solid,mark=square,mark options={solid, mark size=\marksize},line width=\linew] table [ y index={2}, col sep=comma] {../Data/R_2x2_TTI_2_eps_1e-05_T_24_UL.csv}coordinate[pos=0.80](pt1);\addlegendentry{Achievability (\ref{eq:DT})}
\addplot [color=gray, opacity=0.2, forget plot] fill between[of=TTI2_MC and TTI2_DT];

\coordinate (pt2) at ($(pt1) !.5! (ut1)$);
\draw (pt2) ellipse  (3pt and 7pt);
\coordinate (pt3) at ($(pt2)+ (0pt,-6pt)$);
\coordinate (pt4) at ($(pt3)+ (0pt,-10pt)$);
\draw[<-] (pt3)--(pt4) node[anchor=north ] {$\nofdm=2$ ($142.8$ $\mu$s)};
\addplot[name path = TTI4_MC, color=blue,solid,mark=*,mark options={solid, mark size=\marksize},line width=\linew,dashed, forget plot] table [ y index={3}, col sep=comma] {../Data/R_2x2_TTI_4_eps_1e-05_T_48_UL.csv}coordinate[pos=0.80](ut1);
\addplot[name path = TTI4_DT, color=red,solid,mark=square, mark options={solid, mark size=\marksize},line width=\linew, forget plot,dashed] table [y index={2}, col sep=comma] {../Data/R_2x2_TTI_4_eps_1e-05_T_48_UL.csv}coordinate[pos=0.80](pt1);
\addplot [style={pattern=dots,pattern color=gray}] fill between[of=TTI4_MC and TTI4_DT];

\coordinate (pt2) at ($(pt1) !.5! (ut1)$);
\draw (pt2) ellipse  (2pt and 5pt);
\coordinate (pt3) at ($(pt2)+ (0pt,+5pt)$);
\coordinate (pt4) at ($(pt3)+ (0pt,+10pt)$);
\draw[<-] (pt3)--(pt4) node[anchor=south] {$\nofdm=4$ ($285.6$ $\mu$s)};

\addplot[draw=none]{0};

\end{axis}
\begin{axis}
[
	  xmin=1,
	  xmax=25,
      xlabel={Bandwidth [MHz]},
      xticklabels={{$0.18$}, {$0.9$}, {$1.8$}, {$2.7$}, {$3.6$}, {$4.5$}},
      xtick={1,5,10,15,20,25},
      hide y axis,
      axis x line*=top,
      xlabel near ticks
]
    \addplot[draw=none]{0};
    \end{axis}
\end{tikzpicture}
\caption{Achievability bound (\ref{eq:DT}) and converse bound (\ref{eq:MC}) on the maximum coding rate in a UL $2\times 2 $ system for different number of resource blocks $\nres$. Here, $\rho\sub{u} = 20$ dB, $\error = 10^{-5}$, and $\nsubc= 12$. }
\label{fig:UL_2x2}
\end{figure}
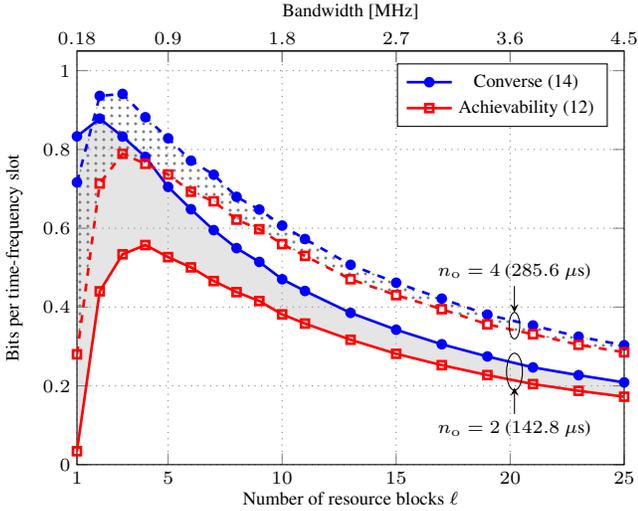

\subsection{Optimal use of spatial and frequency diversity} 
\label{sec:analysis_at_10_9_}

To investigate how to optimally use spatial and frequency diversity, we analyze in this section a \gls{dl} system with a variable number of transmit antennas $\ntx$ and a single receive antenna ($\nrx=1$).
We consider a scenario in which $130$ information bits are transmitted over $\nofdm\nsubc\nres=168$ time-frequency slots ($R\approx 0.77$).
We shall assume $\nofdm=2$ and investigate how the error probability behaves as a function of $\rho\sub{d}$ for different values of $\nres$ and $\ntx$. 
Note that since the total packet length is fixed to $168$, larger values of $\nres$ imply smaller \gls{rb}s.
Specifically, each \gls{ofdm} symbol is assumed to span $\nsubc = 84/\nres$ subcarriers.

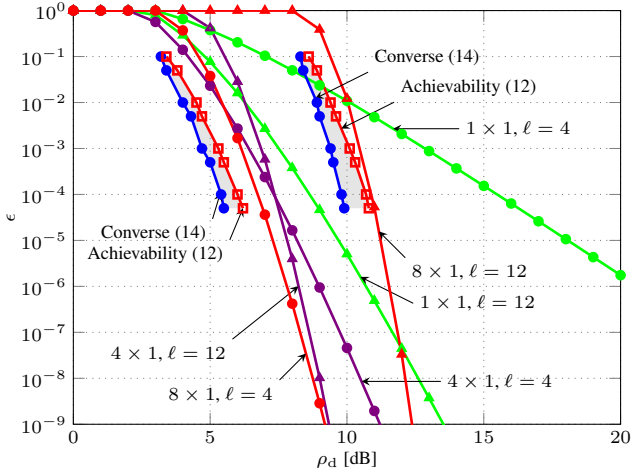
\begin{figure}[t]
\centering
\begin{tikzpicture}
\begin{axis}[%
ymode=log,
xmin=0,
xmax=20,
ymin = 1e-9,
ymax = 1,
ytick={1e0, 1e-1, 1e-2, 1e-3,1e-4,1e-5,1e-6,1e-7,1e-8,1e-9},
yminorticks=true,
xlabel={$\rho\sub{d}$ [dB]},
ylabel={$\epsilon$},
axis background/.style={fill=white},
grid=major,
legend pos= south east
]

\addplot [color = green, solid,mark=*,mark options={solid, mark size=\marksize},line width=\linew] table [y index={1}, col sep=comma] {../Data/EEXP_1x1_L_4.csv}coordinate[pos=0.55](p1);
\coordinate (p2) at ($(p1)+ (25pt,0pt)$); \draw[<-] (p1)--(p2) node[anchor= west] {$1\times 1$, $\nres = 4$};


\addplot [color = green, solid,mark=triangle,mark options={solid, mark size=\marksize},line width=\linew] table [y index={1}, col sep=comma] {../Data/EEXP_1x1_L_12.csv}coordinate[pos=0.41](p1);
\coordinate (p2) at ($(p1)+ (20pt,-10pt)$); \draw[<-] (p1)--(p2) node[anchor= west] {$1\times 1$, $\nres = 12$};


\addplot [color = violet, solid,mark=*,mark options={solid, mark size=\marksize},line width=\linew] table [y index={1}, col sep=comma] {../Data/EEXP_4x1_L_4.csv}coordinate[pos=0.4](p1);
\coordinate (p2) at ($(p1)+ (30pt,0pt)$); \draw[<-] (p1)--(p2) node[anchor=west] {$4\times 1$, $\nres =4$};
3
\addplot [color = violet, solid,mark=triangle,mark options={solid, mark size=\marksize},line width=\linew] table [y index={1}, col sep=comma] {../Data/EEXP_4x1_L_12.csv}coordinate[pos=0.15](p1);
\coordinate (p2) at ($(p1)+ (-25pt,-20pt)$); \draw[<-] (p1)--(p2) node[anchor=north east] {$4\times 1$, $\nres = 12$};

\addplot [color = red, solid,mark=*,,mark options={solid, mark size=\marksize},line width=\linew] table [y index={1}, col sep=comma] {../Data/EEXP_8x1_L_4.csv}coordinate[pos=0.22](p1);
\coordinate (p2) at ($(p1)+ (-10pt,-10pt)$); \draw[<-] (p1)--(p2) node[anchor=north east] {$8\times 1$, $\nres = 4$};

\addplot [color = red, solid,mark=triangle,mark options={solid, mark size=\marksize},line width=\linew] table [y index={1}, col sep=comma] {../Data/EEXP_8x1_L_12.csv}coordinate[pos=0.15](p1);
\coordinate (p2) at ($(p1)+ (10pt,-10pt)$); \draw[<-] (p1)--(p2) node[anchor=north west] {$8\times 1$, $\nres = 12$};

\addplot [name path = MC, color = blue, solid,mark=*,mark options={solid, mark size=\marksize},line width=\linew] table [y index={0}, x index = {1}, col sep=comma] {../Data/SNR_vs_error_8x1_168_12.csv}coordinate[pos=0.28](p1);
\coordinate (p2) at ($(p1)+ (+20pt,+10pt)$); \draw[<-] (p1)--(p2) node[anchor=south west ] {Converse (\ref{eq:MC})};

\addplot [name path = DT, color = red, solid,mark=square,mark options={solid, mark size=\marksize},line width=\linew] table [y index={0},x index = {2},  col sep=comma] {../Data/SNR_vs_error_8x1_168_12.csv}coordinate[pos=0.48](p1);
\coordinate (p2) at ($(p1)+ (+20pt,+10pt)$); \draw[<-] (p1)--(p2) node[anchor=south west ] {Achievability (\ref{eq:DT})};

\addplot [color=gray, opacity=0.2] fill between[of=MC and DT];

\addplot [name path = MC, color = blue, solid,mark=*,mark options={solid, mark size=\marksize},line width=\linew] table [y index={0}, x index = {1}, col sep=comma] {../Data/SNR_vs_error_8x1_168_4.csv}coordinate[pos=0.9](p1);
\coordinate (p2) at ($(p1)+ (-3pt,-10pt)$); \draw[<-] (p1)--(p2) node[anchor=north east ] {Converse (\ref{eq:MC})};
\addplot [name path = DT, color = red, solid,mark=square,mark options={solid, mark size=\marksize},line width=\linew] table [y index={0},x index = {2},  col sep=comma] {../Data/SNR_vs_error_8x1_168_4.csv}coordinate[pos=1](p1);
\coordinate (p2) at ($(p1)+ (-5pt,-12pt)$); \draw[<-] (p1)--(p2) node[anchor=north east ] {Achievability (\ref{eq:DT})};
\addplot [color=gray, opacity=0.2] fill between[of=MC and DT];
\end{axis}
\end{tikzpicture}
\caption{Achievability bound (\ref{eq:eps_eexp}) for a \gls{dl} $\ntx\times 1$ system for different $\ntx$, $\nres$, and $\rho\sub{d}$. It is assumed that $\nofdm=2$, $\nsubc \nres = 84$, and $R=0.77$ bits per time-frequency slot. The achievability bound (\ref{eq:DT}) and converse bound (\ref{eq:MC}) are included for comparison for the cases $\ntx=8$ and $\nres = \lrb{4, 12}$.}
\label{fig:DLeexp}
\end{figure}

In Fig.~\ref{fig:DLeexp}, we plot the achievability bound~(\ref{eq:eps_eexp}) after converting it to maximum error probability~\cite[p. 204]{cover06} for the case $\ntx\in \lrb{1,2,4}$ and $\nres = \lrb{4, 12}$ (which yield $\nsubc = 21$ and $\nsubc=7$, respectively).
For the $8\times 1$ case and both $\nres=4$ and $\nres=12$, we compare the achievability bound~(\ref{eq:eps_eexp}) to the achievability and converse bounds (\ref{eq:DT}) and (\ref{eq:MC}) up to the values of $\epsilon$ for which these two bounds can be computed. 
As expected, (\ref{eq:eps_eexp}) is less accurate than (\ref{eq:DT}) for moderate error probabilities.
For example, when $\ntx = 8$ and $\nres=4$, the gap between these two achievability bounds is $0.26$~dB at $\epsilon=10^{-4}$. The gap turns out to be larger for smaller $\ntx$ values.
For example, when $\ntx=1$, the gap between the two bounds at $\epsilon=10^{-4}$ (not shown in the figure) is $3.8$~dB.

We see from the figure that at $\epsilon=10^{-9}$ the minimum value of SNR $\rho\sub{d}$ predicted by our bound (\ref{eq:eps_eexp}) is achieved by selecting $\ntx=8$ and $\nres=4$, which yields $32$ independent fading branches. 
A similar SNR value is needed when $\ntx=4$ and $\nres=12$, yielding $48$ fading branches. 
The figure also illustrates that further increasing the number of fading branches, as in the $\ntx=8$, $\nres=12$ case, is not effective because one is limited by the channel estimation overhead.
Reducing the number of diversity branches as in the $\ntx=1$, $\nres=12$ case is also not effective, because of the lack of diversity (which is reflected by the more gentle slope of the curves).

\subsection{Practical coding schemes } 
\label{sec:practical_coding_schemes}

We finally benchmark the performance of an actual coding scheme against the bounds provided in Theorems \ref{thm:DT} and \ref{thm:MC}.
We consider a $1 \times 2$ MIMO system in \gls{ul} and assume $\nofdm=2$, $\nsubc=12$, and $\nres=8$, which results in a packet length of $192$ time-frequency slots consisting of $8$ \gls{rb}s.
We also assume that $92$ information bits are transmitted, which results in a rate of $92/192\approx 0.48$ bits per time-frequency slot.
Within each \gls{rb}, we reserve $\np$ time-frequency slots for pilot transmission. 
In the remaining $(\nofdm \nsubc-\np)\nres$ slots we transmit coded bits mapped into QPSK symbols.
As coding scheme, we consider a tail-biting $(368,92)$ convolutional code with a memory-$15$ nonsystematic encoder, which is designed for the case $\np=1$ (indeed, $368$ coded bits yield $184$ QPSK symbols, which together with the $8$ pilot symbols, yield the desired blocklength of $192$).
For values of $\np$ larger than $1$, the encoder output is punctured. Specifically, two coded bits are punctured for each additional pilot symbol.

At the receiver side, the pilot symbols are used to estimate the channel coefficients by means of a maximum-likelihood (ML) estimator.
Thereafter, maximum ratio combining is performed and the bit-wise log-likelihood ratios are derived and given as input to the decoder.
A sub-optimum list decoding algorithm based on ordered statistics has been used for the simulations. 
Specifically, ordered statistics decoding with test patterns of maximum weight equal to $3$ has been adopted. 
This was shown to provide a negligible loss with respect to ML decoding for codes of length up to a few hundred bits \cite{fossorier96}. 
%

\begin{figure}[t]
\centering
\begin{tikzpicture}
\begin{axis}[%
ymode=log,
xmin=16,
xmax=24,
ymin = 1e-3,
ymax = 1,
ytick={1e0, 1e-1, 1e-2, 1e-3,1e-4,1e-5,1e-6,1e-7,1e-8,1e-9},
yminorticks=true,
xlabel={$\rho\sub{u}$ [dB]},
ylabel={$\epsilon$},
axis background/.style={fill=white},
grid=major,
legend pos= south east
]

\addplot [color = purple, solid,mark=*,mark options={solid, mark size=\marksize},line width=\linew] table [x index = {2}, y index={1}, col sep=comma] {../Data/code_performance_np=1.csv}coordinate[pos=0.1](p1);
\coordinate (p2) at ($(p1)+ (10pt,0pt)$); \draw[<-] (p1)--(p2) node[anchor= west] {$\np=1$};

\addplot [color = green, solid,mark=triangle,mark options={solid, mark size=\marksize},line width=\linew] table [x index = {2}, y index={1}, col sep=comma] {../Data/code_performance_np=2.csv}coordinate[pos=0.45](p1);
\coordinate (p2) at ($(p1)+ (10pt,0pt)$); \draw[<-] (p1)--(p2) node[anchor= west] {$\np=2$};

\addplot [color = orange, solid,mark=diamond,mark options={solid, mark size=\marksize},line width=\linew] table [x index = {2}, y index={1}, col sep=comma] {../Data/code_performance_np=4.csv}coordinate[pos=0.6](p1);
\coordinate (p2) at ($(p1)+ (-30pt,0pt)$); \draw[<-] (p1)--(p2) node[anchor= east] {$\np=4$};

\addplot [color = magenta, solid,mark=square,mark options={solid, mark size=\marksize},line width=\linew] table [x index = {2}, y index={1}, col sep=comma] {../Data/code_performance_np=6.csv}coordinate[pos=0.5](p1);
\coordinate (p2) at ($(p1)+ (-10pt,0pt)$); \draw[<-] (p1)--(p2) node[anchor= east] {$\np=6$};

\addplot [color = cyan, solid,mark=star,mark options={solid, mark size=\marksize},line width=\linew] table [x index = {2}, y index={1}, col sep=comma] {../Data/code_performance_np=8.csv}coordinate[pos=0.05](p1);
\coordinate (p2) at ($(p1)+ (-8pt,0pt)$); \draw[<-] (p1)--(p2) node[anchor= east] {$\np=8$};

\addplot[name path = MC, color=blue,solid,mark=*,mark options={solid, mark size=\marksize},line width=\linew] table [y index={0},x index = {3}, col sep=comma] {../Data/DT_MC_code_performance.csv}coordinate[pos=0.8](p1);
\coordinate (p2) at ($(p1)+ (-3pt,-10pt)$); \draw[<-] (p1)--(p2) node[anchor=north east ] {Converse (\ref{eq:MC})};
\addplot[name path = DT, color=red,solid,mark=square, mark options={solid, mark size=\marksize},line width=\linew, forget plot] table [y index={0},x index = {4}, col sep=comma] {../Data/DT_MC_code_performance.csv}coordinate[pos=0.95](p1);
\coordinate (p2) at ($(p1)+ (+9pt, 0pt)$); \draw[<-] (p1)--(p2) node[anchor= west] {Achievability (\ref{eq:DT})};
\addplot [color=gray, opacity=0.2, forget plot] fill between[of=MC and DT];

\end{axis}
\end{tikzpicture}
\caption{Comparison of the achievability (\ref{eq:DT}) and converse bound (\ref{eq:MC}) with the performance of a coding scheme of rate $R\approx 0.48$ bits per time-frequency slot for $\np=\lrb{1,2,4,6,8}$. Here, $\ntx=1$, $\nrx=2$, $\nres=8$, $\nofdm=2$ and $\nsubc=12$.}
\label{fig:codes}
\end{figure}
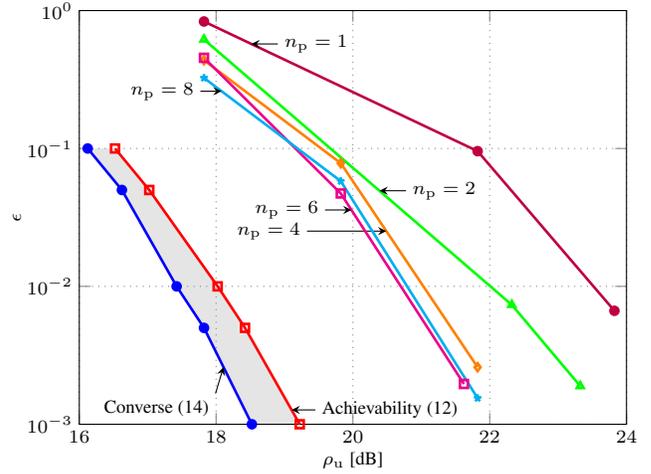

The packet error probability-SNR tradeoff of this coding scheme is depicted in Fig. \ref{fig:codes} for $\np=\lrb{1,2,4,6,8}$. 
As a benchmark, we also depict the performance predicted by the finite-blocklength bounds in Theorem \ref{thm:DT} and \ref{thm:MC}. 
We note that, for the chosen coding scheme, the optimal number of pilots turns out to be $\np =6$.
For this value of $\np$, the SNR gap between the coding scheme and the achievability bound (\ref{eq:DT}) is about $2.68$~dB at $\epsilon=10^{-2}$.
Furthermore, our numerical results illustrate that the performance of our coding scheme is extremely sensitive to the chosen number of pilot symbols. 

\section{Conclusion} 
\label{sec:conclusion}
We considered the problem of designing an OFDM-based system, similar to LTE, operating under stringent constraints on latency and reliability. 
Information-theoretic finite-blocklength bounds turned out to provide valuable insight on how to choose the system bandwidth as a function of the desired reliability and latency constraints, and how to exploit the available spatial and frequency diversity.
We also used our bounds to benchmark the performance of an actual coding scheme, which relies on convolutional encoding  and the transmission of pilot symbols. 

%

 \bibliographystyle{IEEEtran}
 \bibliography{IEEEabrv,confs-jrnls,publishers,references}
 

\end{document}